\documentclass[10pt,conference,a4paper]{IEEEtran}

\usepackage[utf8]{inputenc}

\usepackage{blindtext, graphicx}
\usepackage[utf8]{inputenc}
\usepackage[english]{babel}
\usepackage{csquotes}
\usepackage{wrapfig}
\usepackage{float}
\usepackage{verbatim}
\usepackage{subcaption}
\usepackage{graphicx}
\usepackage{import}
\usepackage{dirtytalk}
\usepackage{subfiles}
\usepackage{array}
\usepackage{multirow}

\usepackage{lscape}
\usepackage{algorithm}
\usepackage{algorithmic}
\usepackage{rotating}
\usepackage[table,xcdraw]{xcolor}

\usepackage{times}

\DeclareGraphicsExtensions{.png,.eps,.ps,.pdf}

\usepackage{url}
\usepackage{hyperref}

\usepackage[most]{tcolorbox}

\usepackage{multirow}
\usepackage{tabularx}
\usepackage{makecell}
\usepackage{svg}

\usepackage{pifont}
\definecolor{ao}{rgb}{0.0, 0.5, 0.0}
\definecolor{amber}{rgb}{1.0, 0.49, 0.0}

\let\labelindent\relax
\usepackage{enumitem}

\usepackage{lscape} 

\graphicspath{{images/}}

\usepackage{amsfonts}

\usepackage{tikz} 
\newcommand{\orcidicon}{\includegraphics[width=0.32cm]{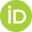}}
\foreach \x in {A, ..., Z}{%
\expandafter\xdef\csname orcid\x\endcsname{\noexpand\href{https://orcid.org/\csname orcidauthor\x\endcsname}{\noexpand\orcidicon}}
}

\begin{document}

\title{Simulating Cyberattacks through a Breach Attack Simulation (BAS) Platform empowered by Security Chaos Engineering (SCE)}

\author{
\IEEEauthorblockN{
\orcidA{}Arturo S\'anchez-Matas$^1$,
\orcidB{}Pablo Escribano Ruiz$^1$,
\orcidC{}Daniel D\'iaz-L\'opez$^{1,2}$\\
\orcidD{}\'Angel Luis Perales G\'omez$^{3}$,
\orcidF{}Pantaleone Nespoli$^{1}$,
\orcidE{}Gregorio Mart\'inez P\'erez$^{1}$
}

\IEEEauthorblockA{$^1$Department of Information and Communications Engineering, University of Murcia, 30100, Murcia, Spain\\
\{arturo.sanchezm, pablo.escribanor, danielorlando.diaz, pantaleone.nespoli, gregorio\}@um.es}
\IEEEauthorblockA{$^2$School of Engineering, Science and Technology, Universidad del Rosario, Bogot\'a, Colombia\\
danielo.diaz@urosario.edu.co}
\IEEEauthorblockA{$^3$Department of Computers Engineering and Technology, University of Murcia, 30100, Murcia, Spain\\ angelluis.perales@um.es}

}

\maketitle

\begin{abstract}
In today’s digital landscape, organizations face constantly evolving cyber threats, making it essential to discover slippery attack vectors through novel techniques like Security Chaos Engineering (SCE), which allows teams to test defenses and identify vulnerabilities effectively. This paper proposes to integrate SCE into Breach Attack Simulation (BAS) platforms, leveraging adversary profiles and abilities from existing threat intelligence databases. This innovative proposal for cyberattack simulation employs a structured architecture composed of three layers: SCE Orchestrator, Connector, and BAS layers. Utilizing MITRE Caldera in the BAS layer, our proposal executes automated attack sequences, creating inferred attack trees from adversary profiles. Our proposal's evaluation illustrates how integrating SCE with BAS can enhance the effectiveness of attack simulations beyond traditional scenarios, and be a useful component of a cyber defense strategy.

\end{abstract}


{\bf Contribution type:}  {\it  Original research }

\section{Introduction}\label{intro}

Cyber defense encompasses the strategies, technologies, and processes designed to protect networks, devices, programs, and data from cyberattacks~\cite{Kizza2020}. It involves implementing security measures such as firewalls, intrusion detection systems, and encryption to safeguard information integrity, confidentiality, and availability. Simulating cyberattacks is crucial in cyber defense as it enables organizations to proactively identify and address vulnerabilities before they can be exploited by adversaries~\cite{Analysis2023}. Through realistic attack simulations, defenders can evaluate the effectiveness of their security controls, improve incident response strategies, and enhance overall resilience against evolving cyber threats. This proactive approach ensures that defense mechanisms are robust and adaptable, ultimately strengthening an organization’s security posture.

Current methods for simulating adversary behavior in cyberspace primarily involve the utilization of automated frameworks and deception techniques to replicate known attack strategies~\cite{Mirage2023}. Tools such as adversary emulation frameworks are specifically designed to mimic the actions of potential attackers and assess the effectiveness of existing security measures~\cite{Smith2022}. Additionally, Honeypots and Threat Intelligence Platforms are crucial in profiling adversary behavior by attracting attackers and providing detailed information about their Tactics, Techniques, and Procedures (TTPs). While Honeypots capture and analyze the actions of real adversaries, Threat Intelligence Platforms aggregate indicators of compromise and attack vectors from current and past campaigns, aiding in the development of accurate adversary profiles. However, traditional simulation methods face significant challenges, including the difficulty of adapting to rapidly evolving threat landscapes and accurately replicating the sophisticated tactics of advanced adversaries. These limitations can result in incomplete assessments of an organization's security posture, leaving critical vulnerabilities undiscovered and unaddressed.

In this scenario, Breach Attack Simulation (BAS) platforms are essential for enhancing cyber defense strategies by enabling organizations to simulate realistic cyberattacks in a controlled environment~\cite{Mohamed2022}. That is, BAS facilitates the creation of comprehensive and repeatable attack scenarios that help security teams evaluate the effectiveness of their defenses, identify vulnerabilities, and improve incident response protocols. Various BAS solutions are available in the market, including proprietary and open-source platforms. Proprietary BAS tools, such as SafeBreach and AttackIQ, offer extensive features and dedicated support, providing organizations with robust capabilities to simulate complex attack vectors. Open-source frameworks such as MITRE Caldera and Atomic Red Team provide adaptability and extensibility, allowing security teams to tailor simulations to their specific needs~\cite{Markus2024}. 

A novel methodology, known as Security Chaos Engineering (SCE), used to evaluate security in systems emerged in the last years and is winning popularity in industry and academy. SCE is a transformative approach that applies the principles of chaos engineering to the security domain. It involves deliberately injecting faults or simulated attacks into a live system to observe how it behaves under adverse conditions. By intentionally triggering failures, SCE aims to expose hidden vulnerabilities and validate the resilience of complex, modern software systems. SCE experiments can range from simulating network congestion to testing the limits of authentication mechanisms, with each test designed to mimic real-world attack scenarios. As a result, SCE transforms a reactive security stance into a proactive and adaptive one, fostering a culture of continuous learning and improvement~\cite{RinehartShortridge}.

The main objective of our research is to enhance organizational defenses by proactively identifying vulnerabilities, thereby enabling continuous improvement in their security posture. To achieve this, we design an innovative framework that utilizes a BAS platform and SCE as integrated means. In this paper, we present the design of our framework, demonstrating how these integrated means facilitate a comprehensive approach to proactive security testing. All of the material developed in our proposal is available as open-source code in the project repository.

The main contributions of this paper are summarized as follows:
\begin{itemize}
\item Proposal of a novel method to simulate adversary behavior through the integration of a BAS platform with a SCE methodology. This seamless combination allows for dynamic and realistic security assessments by leveraging the strengths of both BAS platforms and SCE principles.
\item The composition of Attack Trees (AT) that leverage existing databases of TTPs to facilitate realistic simulations. By utilizing comprehensive TTP databases, our framework constructs detailed and accurate attack scenarios that mirror real-world adversary strategies.
\item The validation of our proposal in a simulated scenario composed by a critical-target information system attacked by an advanced adversary, who is known for having access to a TTP database. This simulation demonstrates that our proposal is reliable and effective in identifying the possible attack vectors in a target systems.
\end{itemize}

This paper is structured as follows: Section~\ref{sota} reviews the most recent works related to BAS and SCE, evaluating their strengths and weaknesses. Next, Section~\ref{proposal} introduces our proposed framework for simulating cyberattacks. In Section~\ref{experiments}, we describe the experiments performed using our framework to validate its effectiveness. Finally, Section~\ref{conclusions} concludes the paper, discussing the implications of our findings and outlining directions for future research.

\section{State of the art}\label{sota}

In this section we will present several research papers related to BAS, attack trees and SCE, which are relevant to understand the novelty of our proposal.

In particular, an attack emulation proposal, named SpecRec, applicable to a variety of steps of a cyberattack was described in~\cite{SpecRep}. It analyzes white papers related to known groups or Advanced Persistent Threats (APT), like Mirai, to extract attack specification that will be emulated later. SpecRep leverages an LLM to translate selected white papers into high-level attack specifications, which are used internally to define the attack's logic. SpecRec also defines a metalanguage to extract attack objectives and a compiler to build attack scenarios. In this way, SpecRep is analog to a BAS platform, where it can generate new attack flows, and is not limited to a predefined playbook of attacks.

Kijong Koo et al.~\cite{AttackGraphGeneration} proposed a novel machine learning-based method to generate attack graphs to predict attack paths. In order to perform this task, two stages were set. The first one revolves around learning the attack paths from known databases relying on machine learning, and the second one consists of generating attack graphs using information such as the network topology and the systems in it. This approach promises to tackle the main issue around the generation of attack graphs, which is the amount of time required to build them manually. 

Basit Ajmal et al.~\cite{ThreatBased} proposed an approach for adversary emulation based on the MITRE ATT\&CK framework to provide an effective way of testing the defense of cyber systems, allowing the profiling of adversaries. Such a paper explored the idea of automated attack generation, facilitating repeated tests and mimicking the behavior of a real-world attack. The simulation developed in such proposal is performed through several means, depending on the attack to be performed, using scripts and adapting payloads.

In a similar way as the previous related work, Sang Ho Oh et al.~\cite{DRL} proposed the use of a Deep Reinforcement Learning (DRL) algorithm to simulate dynamic cyberattacks in scenarios derived from the MITRE ATT\&CK framework. This algorithm is applied to agents that simulate cyberattacks in an environment that mimics a real-world scenario with real vulnerabilities. The DRL agents are trained to adapt on the fly and change their strategies depending on the interactions within the scenario, improving the accuracy of the simulations. 

On the other side, we have ChaosXploit, which is a SCE framework proposed by Palacios et al.~\cite{ChaosXploit}, that facilitates the implementation of the SCE methodology through different experiments. ChaosXploit leverages attack trees to perform experiments attacking cloud infrastructure, such as AWS S3 buckets. This is done thanks to a ``hypothesis generator'', which forms part of a ``knowledge database'', an ``observer'' who has knowledge about the system and the experiments being run, and an ``experiment runner'', in charge of running the experiments.

Another application of SCE was found in~\cite{DevSecOps}. Such a paper studies the security of the traditional DevSecOps lifecycle, arguing that new software vulnerabilities might appear as early as the design phases. It complements the conventional DevSecOps practice through the use of LLMs to automate threat discovery. This proposal also applies the SCE methodology to execute more complex and chaotic tests, which would not be possible with more conventional or standard approaches.\\ 

After analyzing the previous papers, we realized that there was no previous proposal combining both BAS and SCE. BAS platforms suffer from a lack of variety when it comes to choosing the attack vectors to use, and a failure on a chosen vector will often lead to the attack ending. Thus, integrating BAS and SCE sounds like a promising idea that would allow to compose attack trees from the information contained in a threat intelligence database. In addition, cyberattacks would benefit from the complexity and freedom of SCE, without suffering the rigidity of a BAS platform that is limited to a set of attack vectors.

\section{Proposal of Attack Simulations empowered by Security Chaos Engineering}\label{proposal}

\begin{figure*}[ht]
\centering
	\includegraphics[width=0.9\textwidth]{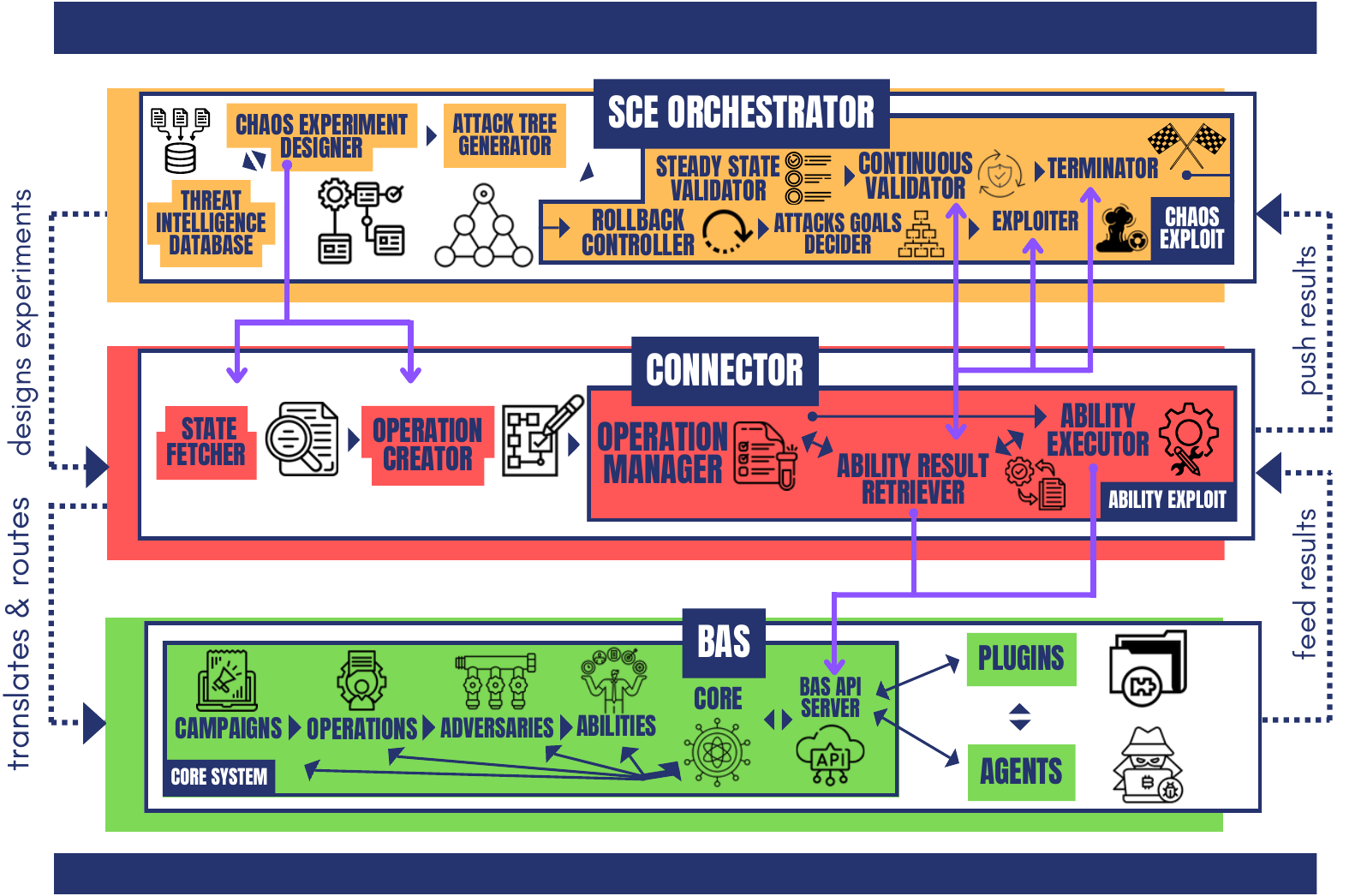} 
	\caption{Three-layered architecture combining BAS with SCE}
	\label{fig:SCEArchitecture}
\end{figure*}

This section presents our proposed solution for integrating SCE into adversary simulation workflows. The architecture of the proposal, illustrated in Figure ~\ref{fig:SCEArchitecture}, is organized into three interconnected layers: i) \textit{BAS layer}, which is based on MITRE Caldera and executes automated attack sequences using real-world adversary profiles, ii) \textit{Connector layer}, which is a middleware layer that interfaces with external platforms and standardizes communication; and iii) \textit{SCE Orchestrator layer}, who is responsible for designing and managing chaos-driven attack scenarios. Together, these 3 layers facilitate systematic and adaptive testing of defenses by combining the unpredictability of SCE with the precision of BAS tools. The following subsections detail the design, functionality, and integration of each component within this modular framework.

\begin{figure}[!th]
\centering
	\includegraphics[width=0.5\textwidth]{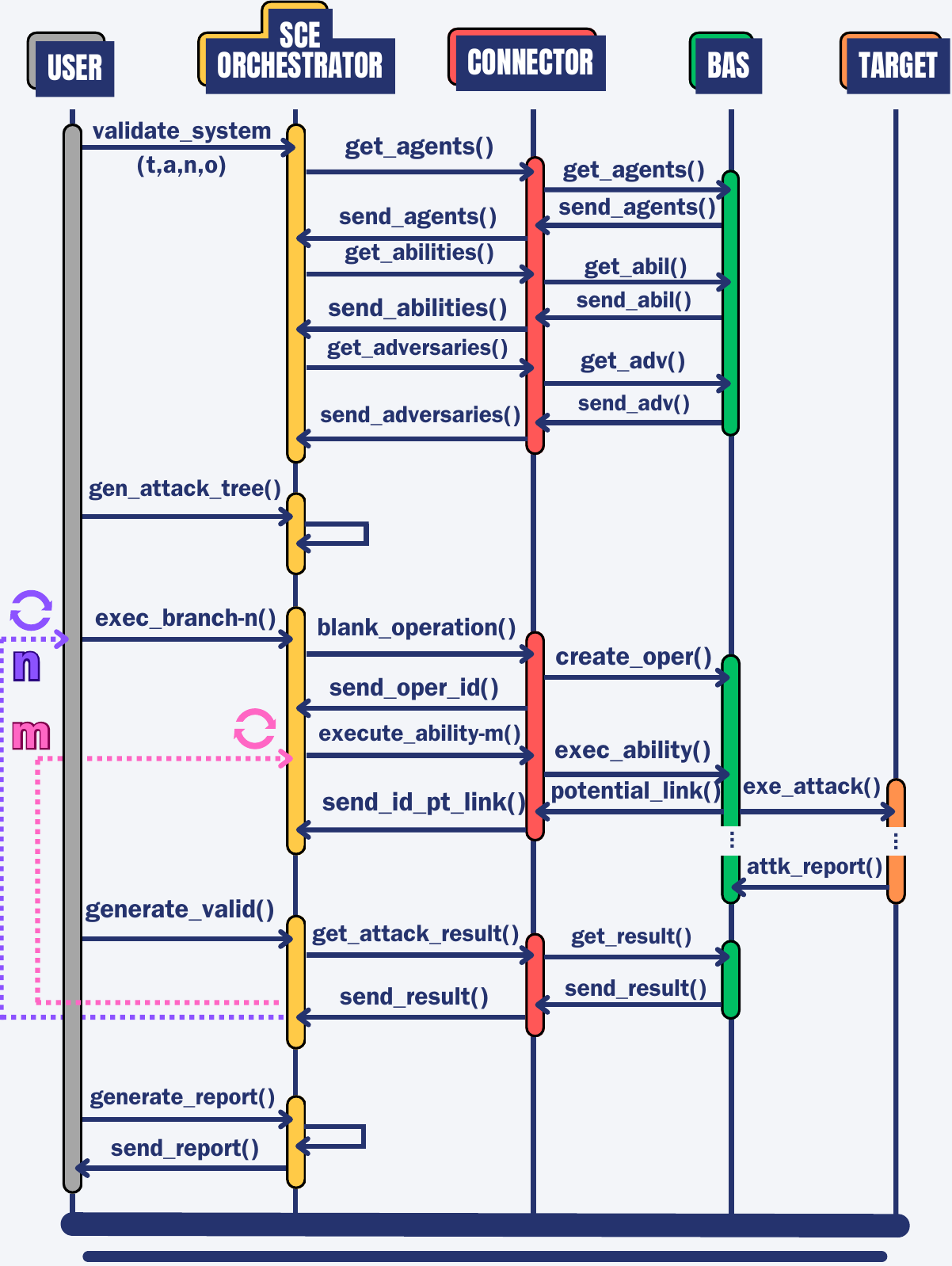} 
	\caption{Flow diagram depicting the interaction between the modules of the proposed architecture}
	\label{fig:FlowDiagram}
\end{figure}

\subsection{BAS layer}\label{BAS}
The BAS layer allows us to run controlled and defined attacks on targets of our choice, giving the possibility of checking what the systems are vulnerable to, and seeing how they were attacked. Our \textit{BAS layer} is implemented with MITRE Caldera, which is an open source BAS solution developed by MITRE. The Caldera module behind the decision-making process was adjusted to allow the SCE methodology to work in conjunction with it, leveraging some native Caldera modules, such as agents, operations, and abilities, which are vital for the proposal and provide the groundwork from which we have expanded.\\

\subsubsection{Core System} The core system, which lays at the foundation of Caldera, allows the management of each of its components, like campaigns, operations, adversaries, and abilities, as well as the intercommunication between each of them. This core system offers an API connection so the \textit{connector layer} can communicate with it via API requests.

\subsubsection{Agents} 
To provide proper attacker simulation, Caldera offers a selection of different agents, i.e. attackers, which are not so different in functionalities from a Remote Access Trojan (RAT). These agents communicate with the Caldera core system, which acts as a Command and Control (C2) server, providing information about the target, reporting result of the actions they perform, and receiving orders about the attacks to perform. Different parameters of the agents can be adjusted to simulate real attackers, like sleep timers, which are useful when trying to avoid detection.

\subsubsection{Abilities} (referred to as Technique in MITRE TTP Matrix) An ability is a term used to describe the action that an agent is going to take, the instruction it is going to receive, and even the attack that it is going to perform; all of the terms can be used interchangeably.

\subsubsection{Adversary profiles} An adversary profile determines which actions will be taken in an attack from start to finish to achieve an attack goal, which could be data ex-filtration, discovery, or many other options. Thus, an adversary profile is comprised of several abilities, and a user can choose to run the abilities he/she deems necessary.

\subsubsection{Operations} An operation is a ``scene'' of sorts, a playground for Caldera to work in. In our proposal, we use a blank operation which does not contain any adversary profile so that the \textit{SCE Orchestrator layer} can decide which actions to execute freely.

\subsubsection{Campaigns} Caldera offers us the possibility of managing a scenario with offensive and defensive actions simultaneously. This scenario is referred to as a campaign, where multiple operations, ones for defense and ones for offense are executed by red and blue agents. Each campaign can mirror real-life scenarios, allowing to test the resilience of an organization via a controlled and repeatable exposure to multiple campaigns.

\subsubsection{Plugins}
Finally, Caldera allows for easy extensibility through plugins which are software components addressed to do specific tasks, e.g. allowing to view the interaction between blue and red agents, simulating human behavior in a target machine, among others.

\subsection{Connector layer}\label{CONNECTOR}
The \textit{Connector layer} is responsible for ensuring communication between the \textit{SCE Orchestrator layer} and the \textit{BAS layer}. Whenever a specific branch of the attack tree is chosen by the \textit{SCE orchestrator layer}, the \textit{connector layer} will execute the abilities associated with this branch, for which it will use a set of API requests addressed to different endpoints available in the \textit{BAS layer}. The \textit{connector layer} is composed of a set of modules described next.\\

\subsubsection{State Fetcher} The state fetcher enables the initial information recollection step of our proposal, allowing us to obtain the general state of the BAS before progressing further. This module permits a swift collection of relevant \textit{BAS layer} information, such as the agents deployed, available abilities, and existing adversary profiles. The collected information mentioned is used to create an ``operation'', which allows the execution of a selected ``adversary profile'' ant its ``abilities'', through an agent deployed in the target.

\subsubsection{Operation creator} All actions in the BAS have to exist in the context of an operation, so this module ensures the creation of a blank operation in which all subsequent decisions can be made, like a blank canvas waiting for a painter.

\subsubsection{Operation Manager} Once the operation is created, there is a certain degree of settings that can be modified before running the experiments. These settings can range from a set of actions, such as modifying the sleep timer of agents, the trusted timer to mark agents as untrustworthy, or, more importantly for the experiments, mutating the operation adding new abilities to execute. These abilities are called potential links, which is a technical term for an ability or technique, and is a way of telling the operation what abilities should be executed.

\subsubsection{Ability executor} After having a blank operation, we can proceed by traversing the attack tree and selecting the next action. For these actions to be performed, the ability executor is in charge of performing the communication with the \textit{BAS layer}'s API to request the execution of an action, i.e., an ``ability''. 

\subsubsection{Ability result retriever} Once an ability is executed, the result is not available immediately. Thus, the ability result retriever ``polls'' the \textit{BAS layer} to validate if a result has been generated, facilitating the  recovery of the output of an attack. The reason for this behavior is that the agents do not perform actions in real time, but according to timers that simulate real threat agents that might want to remain inconspicuous. 

\subsection{SCE Orchestrator layer}\label{SCE_ORCHESTRATOR}
The \textit{SCE Orchestrator layer} is the one devoted for planning and managing chaos-based attack simulations. In particular, it is responsible for accepting user inputs (target machine ($t_i$), agent to use ($a_i$), number of parallel experiments ($n$), and security objective ($o_i$)) to create customized attack scenarios. Leveraging these inputs, it oversees the execution of experiments, ensuring efficient alignment with overall security strategies. This layer manages the entire simulation lifecycle, from setup to teardown, while maintaining consistency across multiple runs. Additionally, it incorporates safety mechanisms to monitor system stability during tests and automatically reverts actions if unexpected issues occur. By transforming security goals into structured chaos experiments, the orchestrator enables teams to evaluate and enhance their defenses under realistic and unpredictable conditions. This layer is composed by a set of modules described next.

\subsubsection{Threat intelligence database} The SCE orchestrator consumes a threat intelligence database to identify the attack procedures which are applicable to a target, according to aspects like the type of operative system, the expected attack goal, the requirements for execution, among others. This database can be one included in the BAS or an external one.

\subsubsection{Chaos Experiment Designer} The Chaos Experiment Designer module defines the scope and parameters of chaos-driven attack simulations. It allows users to specify requirements such as the target machine ($t_i$), deployment agent ($a_i$), number of parallel experiments ($n$), and security objective ($o_i$) (e.g., testing ransomware resilience). By integrating with threat intelligence databases, it dynamically selects adversary Tactics, Techniques, and Procedures (TTPs) that align with the defined goals, ensuring simulations reflect real-world attack patterns.

\subsubsection{Attack Tree Generator}
The attack tree begins with a central attack objective, such as unauthorized access to sensitive data. From this root node, multiple branches extend, each representing a unique pathway an adversary might take to achieve the objective, based on adversary profiles sourced from MITRE Caldera. For example, one branch may involve exploiting software vulnerabilities to gain initial access, while another branch might utilize phishing techniques to obtain user credentials. Each branch further subdivides into specific tactics and techniques, aligning with the principles of Security Chaos Engineering to ensure realistic and unpredictable attack scenarios. This structured approach allows for a comprehensive simulation of diverse attack vectors, facilitating the identification and mitigation of potential security weaknesses within the target architecture.

\subsubsection{ChaosXploit} We use the implementation of ChaosXploit to control the logic of SCE experiments, adapting the components needed, and taking advantage of a previously defined framework~\cite{ChaosXploit}. ChaosXploit is composed of modules, with two of the most important being the Exploiter and Continuous Validator. On the one hand, the Exploiter module executes attacks based on the attack tree structure, ensuring targeted actions within the simulation environment. It utilizes selected adversary tactics to perform precise attack steps, effectively challenging the security posture. On the other hand, the Continuous Validator module continuously monitors system performance and stability throughout the simulations. It tracks real-time operational parameters to detect anomalies and ensures immediate responses are initiated when deviations occur, maintaining the integrity and reliability of the experiments.\\

The flow between the modules of our proposed architecture is depicted in Figure~\ref{fig:FlowDiagram}. The \textit{SCE Orchestrator layer} receives the user inputs (target machine ($t_i$), deployment agent ($a_i$), number of parallel experiments ($n$), and security objective ($o_i$)) and starts the validation. To achieve the validation, the \textit{SCE Orchestrator layer} communicates with the \textit{connector layer}, which also communicates with the \textit{BAS layer} to obtain available agents, abilities, and adversaries. Finally, with all the obtained information (agents, abilities, adversaries), the \textit{SCE Orchestrator layer} checks all the required components to start an attack simulation.

When the user requests the generation of an attack tree \texttt{gen\_attack\_tree()}, the \textit{SCE Orchestrator layer} initiates the Attack Tree Generator module. It analyzes adversary profiles from the Threat Intelligence Database to infer potential attack paths. By leveraging predefined tactics and techniques, the generator constructs a structured attack tree, detailing each possible step based on the input parameters. The completed attack tree is then ready for execution, offering a comprehensive visual of potential threat vectors.

Upon the user's request to execute a branch \texttt{execute\_branch-n()}, the \textit{SCE Orchestrator layer} coordinates with the \textit{BAS layer} to establish an operation. This involves executing ``m'' abilities designed within the attack tree branch. Each ability is processed sequentially to reflect real-world attack scenarios. The execution may involve ``n'' branches, depending on the complexity and objectives defined by the user. Throughout this process, the \textit{BAS layer}'s Core Framework manages the Attack Execution Engine to ensure precise application of the designed sequences. After execution, the results are gathered and prepared for analysis in the form of an attack report \texttt{attack\_report()}.

After completing the execution, the \textit{SCE Orchestrator layer} retrieves the attack results using \texttt{get\_attack\_result()}. It systematically collects the outcomes and performs an initial analysis of the simulation's effectiveness. The processed results are then sent back to the user \texttt{send\_result()} to validate the attack's success and effectiveness, providing insights into potential security improvements.

When the user requests to generate a report \texttt{generate\_report()}, the \textit{SCE Orchestrator layer} compiles all data and findings into a comprehensive document. This report encapsulates the simulation process, including objectives, methodologies, executed abilities, and results. Moreover, such a report highlights key vulnerabilities discovered and suggests potential remediation steps. This detailed report serves as a critical tool for organizations to understand their current security posture and prepare strategies for future improvements.

\section{Experiments}\label{experiments}

Several experiments were created, executed, and reviewed for the validation of our proposal. All experiments are available in the project repository\footnote{\url{https://github.com/noname13a/SCE-BAS}}. To ensure repeatability, Section~\ref{subsec:IdsSettings} describes the scenario, where the hardware and software elements will be detailed,  Section~\ref{subsec:Parameters} describes the parameters needed for the simulation, Section~\ref{subsec:AttackTree} indicates the attack tree that will be followed by the simulated adversary,  Section~\ref{subsec:SCEexperiment} defines the SCE experiment, and
Section~\ref{subsec:Simulation} resumes the results of execution.

\begin{figure}[h]
    \centering
    \includegraphics[width=0.9\columnwidth]{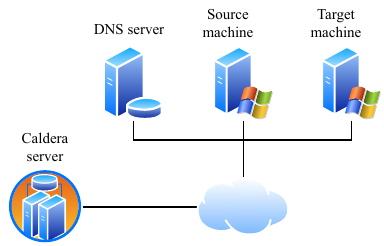} 
        \caption{Network topology for the experiments}
        \label{fig:Network infrastructure}
\end{figure}

\begin{figure*}[!th]
\centering
    \includegraphics[width=0.7\textwidth]{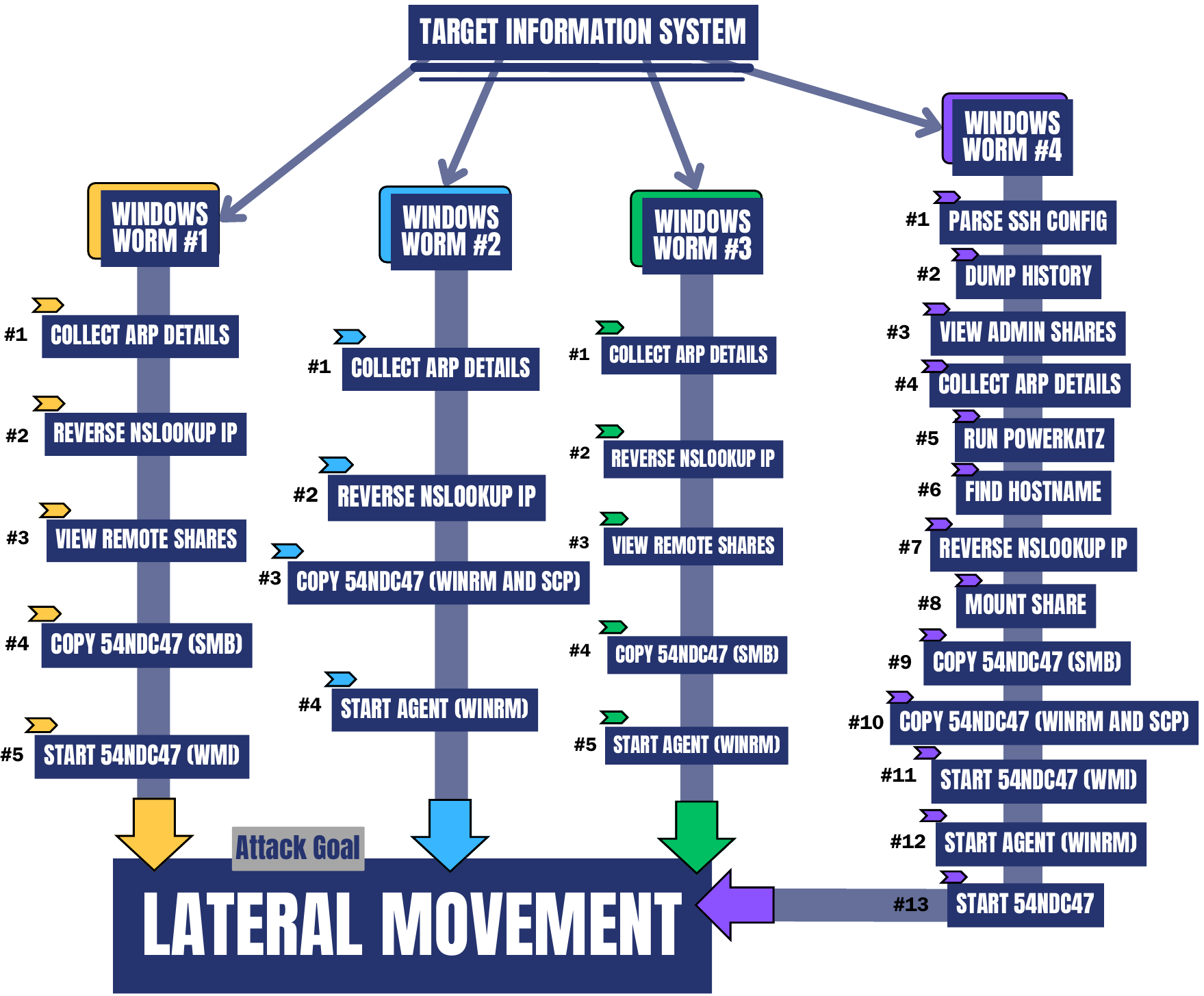} 
    \caption{Attack Tree composed from adversary profiles existing in Caldera}
    \label{fig:attacktree}
\end{figure*}

\subsection{Settings}\label{subsec:IdsSettings}

The components required for the experiments were virtualized through a host machine with an Arch-based distribution, AMD Ryzen 7 5800X CPU, Nvidia RTX 4070 GPU, 32GB RAM, and 2TB SSD. Four virtual machines were used in the experiments as detailed next, and following the network topology of Figure~\ref{fig:Network infrastructure}:

\begin{itemize}   
    \item Caldera server: This machine serves as the location where the BAS will be located and running. It is the central point where the agents report to, and from where they receive the instructions. Our proposal does not have to be run on the same machine, but on a machine which can access its API through the network. Caldera version 4.2.0. was virtualized on an Ubuntu Desktop 22.04 distribution with stock settings.
    \item Source machine: The source machine serves as the initial point for the cyberattack, being a machine infected by a worm that is trying to perform lateral movement to other targets within the network. It is virtualized on a Windows 10 base installation. 
    \item Target machine: The target machine will represent our victim, which is a machine that shares a network with the source machine. It has some settings that make it vulnerable, such as a vulnerable SMB server configuration, and weak security policies. 
    \item DNS server: A DNS server which will allow both the target and source machine to have a fully qualified domain name (FQDN), albeit a local one. This is a requirement to run certain adversary profiles and allows the communication to be done with natural names instead of IP addresses.
\end{itemize}

\subsection{Defining the parameters of the simulation}\label{subsec:Parameters}
As previously mentioned, some variables have to be determined in preparation for the simulation. First, it would be the target machine ($t_i$) chosen between a list of available targets (($TargetList=\{t_1, ..., t_n\}$)). Without a target, other machines found on the network could be attacked, which may or may not be the expected behavior. Setting this target ensures the attack is directed at the machine we want to test. Secondly, we must specify a deployed agent ($a_i$) existing in an agent list ($AgentList=\{a_1, ..., a_n\}$), which will be used to execute the abilities determined on the attack graph. Though Caldera offers us the possibility of using several agents on an operation, each with a different role, for our experiments we will limit ourselves to one agent, trying to pivot to the target machine. The last two parameters are the number of parallel experiments ($n$), which is a default value, and the security objective ($o_i$), will be chosen between a list of possible available objectives ($ObjectiveList=\{o_1, ..., o_n\}$). The selected variable values for our experiments are described in Table~\ref{tab:variables}.

\begin{table}[]
\resizebox{\columnwidth}{!}{%
\begin{tabular}{|c|c|c|}
\hline
\textbf{\begin{tabular}[c]{@{}c@{}}Variable \\ name\end{tabular}} & \textbf{Variable value(s)} & \textbf{\begin{tabular}[c]{@{}c@{}}Selected value\\ for the experiment\end{tabular}} \\ \hline
$t_i$ & \begin{tabular}[c]{@{}c@{}}$TargetList=\{Windows10_A,$\\ $WindowsServer_A, UbuntuServer_A\}$\end{tabular} & $t_1 = Windows10_A$ \\ \hline
$a_i$ & \begin{tabular}[c]{@{}c@{}}$AgentList=\{sandcat_A$, \\ $sandcat_B, sandcat_C\}$\end{tabular} & $a_1=sandcat_A$ \\ \hline
$n$ & $n \in \mathbb{N}^+$ & $1$ \\ \hline
$o_i$ & \begin{tabular}[c]{@{}c@{}}$ObjectiveList=\{LateralMovement$, \\ $Exfiltration$, $PrivilegeEscalataion\}$\end{tabular} & $o_1 = LateralMovement$ \\ \hline
\end{tabular}%
}
\caption{Selected variables for our experiment}
\label{tab:variables}
\end{table}

\subsection{Identifying the Adversary Options}\label{subsec:AttackTree}

Our framework composes an attack tree (Figure~\ref{fig:attacktree}) autonomously, extracting attack goals from Caldera's adversary profiles, presented as a comprehensive final report, which serves as a foundational tool for conducting Security Chaos Engineering (SCE) experiments. 

Each branch of the attack tree corresponds to one of these adversary profiles with their respective abilities depicted as nodes within each branch. This structure allows the SCE Orchestrator to utilize the attack tree in conjunction with the SCE philosophy to design chaos-driven experiments that emulate realistic adversary behaviors. For this example, we have filtered the attack tree to include adversary profiles with common attack goals of discovery, lateral movement, and execution, although some profiles possess additional abilities. This attack tree integrates multiple adversary profiles, each sharing common attack objectives such as discovery, lateral movement, and execution. Specifically, in our experiment, we have included four distinct adversary profiles: Windows Worm \#1 (SMB + WMI), Windows Worm \#2 (WinRM + SCP), Windows Worm \#3 (SMB + WinRM), and Worm (SMB + WinRM + WMI).

The worms described utilize various techniques for lateral movement, with some overlapping techniques. Those worms typically begin by collecting ARP details and reversing nslookup IPs to map the network environment. The movement is achieved by using protocols like SMB, WinRM, and SCP to copy and start processes on remote machines. For instance, Windows Worm \#1 utilizes SMB and WMI to move laterally, while Windows Worm \#2 employs WinRM and SCP. Windows Worm \#3 combines both SMB for copying and WinRM for execution. The general approach involves viewing remote or admin shares, copying essential files, and starting agent processes to ensure the worm propagates effectively. The Worm adversary example illustrates a combination of these techniques, enhancing its lateral movement capabilities by parsing SSH configurations and utilizing both WMI and WinRM for process initiation, thus ensuring versatility and robustness in achieving its objectives across a network.

\subsection{Composing SCE experiments}\label{subsec:SCEexperiment}
Our goal with this experiment stems from the question of how secure would an organization be in front of a self-replicating worm giving attackers Command and Control (C2) and Remote Access Tool (RAT) control of the computers in the network. This experiment could be performed by an organization interested in simulate cyber attacks coming from an adversary with a known arsenal, with the purpose of identify which attack vector, i.e. a branch of the tree, could be successful. Based on the attack tree presented beforehand, we can define the experiment following the SCE scientific method the following way:
\begin{itemize}
    \item \textbf{Observability}: We have several options for observability, all related to the creation of a new agent. First, the response from Caldera for the ability execution, which we can parse to determine if it has been successful. Another would be obtaining this information through API calls, in particular, \texttt{getAgents()}, which could be used to compare it to the base call result, seeing if we have a new agent on the list.
    \item \textbf{Steady State}: The system in its base state, without running experiments, representing an objective that has not been attacked yet.
    \item \textbf{Hypothesis}\label{Hypothesis}: Assuming proper configuration of the services and security of the system, a lateral movement attack, like a self-replicating worm, should fail.
\end{itemize}

\subsection{Execution of simulations}\label{subsec:Simulation}
Following the execution of our experiments, we will explain first the branches that succeeded, being Windows Worm \#1 and \#3. Those reached the end, refuting the hypothesis mentioned in Section~\ref{subsec:SCEexperiment}, and proving that the worm was allowed to pivot to a different machine within the network.\\

Both worm \#1 and \#3 managed to implant the agent on the source machine through a shared SMB folder with incorrect permissions assigned, allowing unauthenticated users to copy files to the public directory of the machine. This file was then executed, via different methods, frist with WMIC and then with WinRM, both succeeding due to weak credentials (Worm \#1) and improper configuration (Worm \#3).\\

As for the branches that failed, starting with Windows Worm \#2, this branch aborted due to not having the knowledge required to propagate the worm through SCP, this is the reason behind the failure of branch \#4 as well. This failure could be mitigated with intermediate steps, which could try to bypass or find the credentials of the target machine, allowing its implantation. Once implanted, the execution methods are the same as the branches that succeeded, meaning that the only reason those branches failed was the lack of credentials. Details about the execution of experiments and the obtained results can be consulted in the project repository.

\section{Conclusions and Future Work}\label{conclusions} 

In this paper, we introduce a framework that integrates Breach Attack Simulation (BAS) tools with Security Chaos Engineering (SCE) methodologies to conduct comprehensive security experiments across diverse target architectures. Leveraging hypotheses generated from our knowledge database and detailed attack representations, our proposal executes SCE-driven experiments to identify potential security vulnerabilities within target systems. To demonstrate the effectiveness of our proposal, we conducted a series of experiments a cyber defense scenario, assessing the security posture through SCE-based simulations. The results highlighted that some adversary profiles, i.e. some branches of the attack tree, successfully exploited the target. Our proposal is publicly available to the cybersecurity community through our the project repository\footnote{\url{https://github.com/noname13a/SCE-BAS}}.

Even with the results of the experiments, the proposal is still far from complete, as it requires an extensive amount of tinkering and work, and the range of scenarios it covers is fairly limited. We intend to extend the functionality to work with all of the adversary profiles provided by Caldera, but more importantly, adding the possibility of jumping from one branch to another inside the attack graph, or even taking leaps and backtracking to different nodes in it, expanding the capabilities of the solution developed.

\section*{Acknowledgment}
This work has been co-funded by the European Union (project ECYSAP EYE). Views and opinions expressed are however those of the author(s) only and do not necessarily reflect those of the European Union or the European Defence Fund. Neither the European Union nor the granting authority can be held responsible for them.\\

This work has also been partially supported by MCIN/AEI/10.13039/501100011033 NextGeneration EU/PRTR, UE, under Grant TED2021-129300B-I00, by MCIN/AEI/10.13039/501100011033/FEDER, UE, under Grant PID2021-122466OB-I00, by  the Spanish National Institute of Cybersecurity (INCIBE) by the Recovery, Transformation and Resilience Plan, Next Generation EU under the strategic project DEFENDER, by the CyberDataLab (Cybersecurity and Data Science Laboratory) at the University of Murcia (Spain), and the School of Engineering, Science and Technology at the University of Rosario (Colombia).

\bibliographystyle{IEEEtran}
\bibliography{bibliography.bib}{}

\end{document}